\documentstyle[12pt,aaspp4]{article}

\def\dim#1{\mbox{\,#1}}
\def\tablenumpar{
\begin{table}
\label{tabnumpar}
\caption{Numerical Parameters}
\medskip
$$
\begin{tabular}{ccccccc}
Run & $\Omega_b$ & $N$ & Box size & Total mass res.\
 & Spatial res.\ & Dyn.\ range \\ \tableline
A & 0.055 & $128^3$ & $3h^{-1}{\rm\,Mpc}$  & $10^{6.1}h^{-1}\dim{M}_{\sun}$ & $1.5h^{-1}{\rm\,kpc}$ & $2000$ \\
B & 0.055 & $ 64^3$ & $3h^{-1}{\rm\,Mpc}$  & $10^{7.0}h^{-1}\dim{M}_{\sun}$ & $4.5h^{-1}{\rm\,kpc}$ & $640$ \\
C & 0.055 & $ 64^3$ & $3h^{-1}{\rm\,Mpc}$  & $10^{7.0}h^{-1}\dim{M}_{\sun}$ & $4.5h^{-1}{\rm\,kpc}$ & $640$ \\
D & 0.03  & $ 64^3$ & $2h^{-1}{\rm\,Mpc}$  & $10^{6.5}h^{-1}\dim{M}_{\sun}$ & $3.0h^{-1}{\rm\,kpc}$ & $640$ \\
E & 0.055 & $ 64^3$ & $32h^{-1}{\rm\,Mpc}$ & $10^{10.1}h^{-1}\dim{M}_{\sun}$ & $50h^{-1}{\rm\,kpc}$ & $640$ \\
F & 0.055 & $ 64^3$ & $64h^{-1}{\rm\,Mpc}$ & $10^{11.0}h^{-1}\dim{M}_{\sun}$ & $100h^{-1}{\rm\,kpc}$ & $640$ \\

\end{tabular}
$$
\end{table}
}

\begin{document}
\title{Spectral Features from the Reionization Epoch}
\author{Edward A. Baltz}
\affil{Department of Physics, University of California, Berkeley, CA 94720}
\author{Nickolay Y. Gnedin}
\affil{Department of Astronomy, University of California, Berkeley, CA 94720}
\and
\author{Joseph Silk}
\affil{Departments of Astronomy and Physics, and Center for Particle
Astrophysics, University of California, Berkeley, CA 94720}

\begin{abstract}
Emission lines in hydrogen can be used to measure the approximate redshift of
the reionization of the universe.  This is an important measurement given the
lack of a convincing theoretical prediction of this epoch.  There is a rapid
change in absorption at the reionization event, which can leave sharp features
in the spectrum of recombining hydrogen.  These features in the near infrared
background permit us to measure the redshift at which reionization occurs.  We
have calculated these signals based on numerical simulations of the
$\Lambda$CDM cosmogony.
\end{abstract}


\section{Introduction}

It is well known that the Lyman alpha forest shows very little Gunn-Peterson
absorption along lines of sight to distant quasars.  The standard
interpretation of these observations is that the intergalactic medium (IGM)
becomes highly ionized before redshift $z=5$.  The approximate reionization
redshift is among the most important unknown quantities relevant to large scale
structure and cosmology.  There have been many attempts to simulate the
reionization process, but they are strongly dependent on the input physics and
are hence unreliable.  In fact, it is still quite plausible to have
reionization occur anywhere between redshifts $z=5$ and $z=50$ in most
reasonable cosmological models.  It is observations that constrain the
reionization epoch to this range: the lack of the Compton $y$-distortion of the
CMB excludes $z>50$ and the lack of Gunn-Peterson absorption excludes $z<5$.

Because of the uncertainty in the reionization epoch, it is important to be
able to observe the reionization event in some way.  We propose a new signal in
the cosmic infrared background (CIB) that is potentially capable of giving the
redshift of reionization.  While the signal is small relative to the current
limits at 1-10 $\mu$m, the predicted frequency dependence makes detection
possible.

The rate of hydrogen recombinations is peaked at the reionization epoch, and
this is the source of the background we seek.  The fraction of neutral hydrogen
undergoes a drop of a factor of at least 10$^3$ in about a tenth of a Hubble
time, thus the Gunn-Peterson optical depth undergoes a drastic change in a very
small redshift interval.  This abrupt change in absorption can give an accurate
value for the redshift of reionization.  Because we observe an all-sky
background, we are only concerned with the effects absorption has on the
spectrum.  It serves to brighten some lines while dimming others.  Because the
effect turns off very rapidly, sharp features are produced in the spectrum.

\section{Simulations}

In order to be specific, we adopt a CDM+$\Lambda$ cosmological model as a
framework for our investigations. We fix the cosmological parameters as
follows: $\Omega_0=0.35$, $\Omega_\Lambda=0.65$, and $h=0.70$. We study two
values of the baryon density parameter, $\Omega_b=0.03$ and 0.055.  We allow
for a small tilt in the primordial spectrum of $n=0.96$, and we normalize the
model to COBE according to Bunn \& White (1997).  In order to accurately
determine the initial conditions, we compute the linear transfer functions for
the model using the linear Boltzmann code LINGER from the COSMICS package
(Bertschinger 1995).

The simulations were performed with the SLH-P$^3$M cosmological hydrodynamic
code (Gnedin 1995; Gnedin \& Bertschinger 1996).  The physical modeling
incorporated in the code is described in Gnedin \& Ostriker (1997), and
includes dynamics of the dark matter and cosmic gas, evolution of the spatially
averaged UV- and X-ray background radiation, star formation, stellar feedback,
non-equilibrium ionization and thermal evolution of primeval plasma, molecular
hydrogen chemistry, equilibrium metal cooling, and self-shielding of the gas.

We have performed six runs with different box sizes and numerical resolutions
to assess the importance of different scales and estimate the uncertainty due
to the finite resolution of our simulations as well as to test different
physical assumptions included in the simulations.

The adopted parameters of these runs are compiled in Table \ref{tabnumpar}. The
first four runs were stopped at $z=4$, since at that moment the rms density
fluctuation at the box scale reaches 0.3-0.4, and the absence of fluctuations
with wavelengths larger than the box size renders the simulations
unreliable. The two runs E and F were continued until $z=0$, and are used to
assess the evolution of the recombination rate at low redshifts. Because of
their low resolution, they cannot be used for calculating the reionization
history of the universe.

The largest of our simulations, run A, has the highest resolution, and is our
fiducial run. Its spatial and mass resolutions are sufficient to accurately
predict the epoch of reionization and the evolution of the recombination rate
throughout the reionization period. Two smaller runs, B and C, differ by the
shape of the spectrum emitted by the sources of ionization. Specifically, run C
includes only young stars as sources of ionization, whereas run B includes
quasars in the ionizing radiation.  Finally, run D has a lower value of the
baryon density.  This run is discussed in detail in Gnedin \& Ostriker (1997).

\section{Radiative Transfer}
The space averaged equation of radiative transfer in an expanding universe is 
\begin{equation}
\frac{\partial\overline{J}_\nu}{\partial t}-
H\left(\nu\frac{\partial\overline{J}_\nu}{\partial\nu}-
3\overline{J}_\nu\right)=-\overline{\kappa}_\nu\overline{J}_\nu+
\overline{S}_\nu,
\end{equation}
where $\overline{\kappa}_\nu=\langle\kappa_\nu J_\nu\rangle/\overline{J}_\nu$.
We need the intensity of a sharp line at frequency $\nu_i$ as a function of its
redshift.  The source for an infinitely sharp line is given by
\begin{equation}
\overline{S}_\nu=\frac{h\nu_ic}{4\pi}R\delta(\nu-\nu_i),
\end{equation}
where $R$ is the volume rate of photon emission and $\delta$ is the Dirac delta
function.  Neglecting absorption, the intensity today in a flat
$\Omega_0+\Omega_\Lambda=1$ universe is given by
\begin{equation}
\overline{J}_\nu=4.878\,(\Omega_0h^2)^{-1/2}\left(\frac{\nu}{\nu_i}
\right)^{9/2}\left[1+(\Omega_0^{-1}-1)\left(\frac{\nu}{\nu_i}\right)^3\right]
^{-1/2}R(t_i).
\label{jnu}
\end{equation}
where $t_i$ is defined by $a(t_i)=\nu/\nu_i$ and $a$ is the scale factor
normalized to unity today.

\section{Hydrogen Recombinations}

The case A recombination rate is computed from the simulation data by obtaining
the average $R=\langle n_e n_{H^+}\alpha_A\rangle$ over the simulation box.
The fit to $\alpha_A$ from Hui and Gnedin (1997) was used.  This quantity drops
steadily before reionization as the universe expands.  There is a rise at the
reionization epoch as the hydrogen becomes nearly completely ionized.  After
reionization, the rate declines steadily until the end of the simulation.

From the case A recombination rate we need to obtain the direct recombination
rates to the individual levels.  We use the rates in Ferland et al. (1992) for
the mean temperature of the gas over the simulation box, which is about 10$^4$
K.  For levels $n\ge 7$ we have used the following fit which reproduces the
summed values well at least to $n=20$.
\begin{equation}
\alpha_n\approx -(1.20n^{-2}-38.8n^{-2.5}+56.1n^{-3})\times 10^{-13}
{\rm cm}^3\,{\rm s}^{-1}.
\end{equation}
These three values correspond to the three summed recombination rates at 10$^4$
K in Ferland et al. (1992) for $n=7-10$, $n=11-20$, and $n=21-\infty$.  These
values reproduce the $n=6$ recombination coefficient well, but the $n=6$ rate
was not used in making the fit.  At large $n$ this fit becomes negative, but
this does not happen until $n=1000$ or so.

We use the approximation given by Johnson (1971) for the radiative transition
rates in hydrogen.  Knowing the rates of direct recombinations to various
levels, we can then compute the strengths of lines.  We find that the emission
is dominated by the direct recombination photons.

\section{Absorption}

We have used a simple approximation to account for the absorption.  Because we
are investigating a diffuse signal, we are only concerned with how energy is
redistributed among different spectral lines.  We assume that the optical depth
is very large before reionization and very small afterwards.  This is a
reasonable approximation because the simulations indicate that the neutral
hydrogen fraction drops by three orders of magnitude in about a tenth of a
Hubble time.  We refer to this as the reionization event.

The justification of our approximation is as follows.  The Gunn-Peterson
optical depth is given by $\tau=4.2\times 10^{10} n_I (\Omega_0h^2)^{-1/2}
(1+z)^{-3/2}$.  The current observational limit indicates that $n_I<2\times
10^{-12} (\Omega_0h^2)^{1/2}(1+z)^{3/2}$, giving $\tau\approx 0.1$.  We neglect
this relatively small optical depth.  Before reionization however, the neutral
fraction is a factor of at least 10$^{3}$ higher, possibly as much as a factor
of $10^5$ higher, giving a Gunn-Peterson optical depth of at least $\tau=10^2$,
and probably 10$^4$.  The other lines in the Lyman series will have smaller
optical depths due to their smaller cross sections, but this effect does not
refute this argument.  For example, the optical depth of Ly-7 is only decreased
by a factor of 100.  This justifies our approximation.

We assume that any transition that is not in the Lyman series is optically
thin.  There are a large number of neutral hydrogen atoms for each photon, so
the hydrogens will be virtually all in the ground state.  Each time a Lyman
series transition is excited, there is some chance that it will decay through
other states. These photons can escape unless they are also in the Lyman
series.  The net effect of this is that photons are removed from the Lyman
series and broken up into lower energy photons and Ly$\alpha$.

We now estimate how many times this process occurs.  The probability
to remain in the Lyman series is simply $P=e^{-\tau}$, and $P=P_1^N$, where
$P_1$ is the probability of remaining after one interaction.  The branching
ratios for the $n-1$ transitions are all roughly 95\%, with the $n-2$
transitions making up most of the remaining 5\%, thus $P_1\approx 0.95$, giving
$N\approx 20\tau$.  Through the reionization event, $\tau$ drops from about
$10^4$ to about 1, so previous to reionization, Lyman series photons suffer as
many as $10^5$ absorptions.

We approximate this process by recomputing the branching ratios, taking into
account the fact that radiative transitions in the Lyman series are forbidden,
except for Ly$\alpha$, where there is no other possibility.  These modified
branching ratios are used until the reionization event.  After even 100 such
decays, the probability that every single one was a Lyman series decay is about
$0.5\%$.  Our approximation for the branching ratios is thus justified.  We
find that Ly$\alpha$ becomes about three times brighter and that H$\alpha$
becomes about thirty times brighter.

We neglect opacity due to dust.  Dust is not associated with the highly ionized
gas we are studying, and in any event it is unlikely that there is any
significant dust formation at the high redshifts we are concerned with.  For
example, the average metallicities in the Lyman-alpha forest at redshift $z=3$
are not much in excess of 0.01 solar (Songaila \& Cowie 1996).  More recent
dust formation may obscure a CIB, but we expect that the dust covers only a
small fraction of the sky, so we neglect dust absorption in our calculation.
In any event, dust absorption can not erase a spectral feature, though it may
decrease the total amplitude.

\section{Results}

We have attempted to extrapolate the emission rate to $z=0$ in several cases.
As a simple exercise, we first neglect temperature evolution, thus fixing
$\alpha$.  If the bulk of the emission comes from bound objects, then $R\sim
a^{-3}$.  If the emission is coming from growing mode perturbations in the
linear regime, then $R\sim a^{-4}$.  Finally, if the emission is coming from a
uniform background, then $R\sim a^{-6}$.

We use runs E and F to investigate the low redshift evolution.  At late times,
after reionization, the case A recombination rate varied as roughly $R\sim
a^{-7}$ in these simulations.  The temperature in the predominant emission
regions is thus rising.  We use this result in extrapolating the higher
resolution simulations to $z=0$.  The emission will be suppressed in the
highest density regions due to the greatly increased temperature and thus
greatly decreased recombination rate.

We can make a simple argument to motivate this scale factor dependence.  The
overall temperature of the IGM is around 10$^4$ K.  The recombination
coefficient varies roughly as $\alpha_A\sim T^{-1}$ at temperatures of 10$^6$
K.  The temperature in a cluster can be 10$^7$ or 10$^8$ K.  If a cluster is
overdense by a factor of 10$^3$, the square of the density is up by 10$^6$.
The temperature is up by a factor of roughly 10$^3$, and the total volume of
gas is down by a factor of $10^3$.  Crudely, the cluster produces as much
emission as the volume of IGM from which it formed.  If we assume unresolved
clusters dominate the emission, we can then assign a power law evolution to the
mean temperature of the IGM today, which is as large as 10$^8$ K.  At redshift
$z=10$ in the simulations, the temperature is 10$^4$ K and today it is 10$^8$
K, This means that the temperature rises like $T\sim a^4$, so $\alpha\sim
a^{-4}$, and for clusters, which are bound objects, the total rate varies as
$R\sim a^{-7}$.

There are two quantities that we track at the reionization epoch, the space
averaged recombination rate and the neutral hydrogen fraction.  The hydrogen
becomes ionized in roughly a tenth of a Hubble time at a redshift of about
$z=7$, depending on the model.  The recombination rate is at a maximum at this
time, but the peak is quite broad.  To illustrate this point, we have plotted
the recombination rate and the neutral hydrogen fraction together in figure~ 1
for simulation A.  The other simulations have similar behavior.

In figure~ 2 we plot $\overline{J}_\nu$ from recombinant emission from the
first 20 levels of hydrogen for simulations A-D.  The recombination rate has
been extrapolated in the high resolution simulations from $z=z_{\rm min}$ to
$z=0$ according to a power law,
\begin{equation}
R(z<z_{\rm min})=R(z_{\rm min})\left(\frac{1+z_{\rm min}}{1+z}\right)^
{-\beta},
\end{equation}
where we have used $\beta=7$, the fit obtained from simulations E and F.
Inspecting equation (\ref{jnu}) we see that $R$ must fall more quickly than
about $a^{-5.5}$ at late times in the flat model being considered.  If $R$
does not fall quickly enough, the spectrum is dominated by the most recent
emission, and any feature from the reionization epoch is obscured.

It is the absorption that gives the best indication of the redshift of the
reionization event.  Before reionization, the optical depth in the Lyman series
is very high.  All Lyman series lines except Ly$\alpha$ are absorbed
immediately and redistributed.  This makes Ly$\alpha$ and the Balmer series
significantly brighter because they receive the energy of the Lyman series.
Once reionization occurs, the Lyman series becomes optically thin and the
brightness of Ly$\alpha$ and the Balmer series drops considerably.  This is
clearly seen in the spectrum.  As we look redward, crossing Ly$\alpha$ at about
one micron or H$\alpha$ at about six microns from the reionization event, the
brightness suddenly increases due to the effect of the absorption and
redistribution of the higher Lyman series lines.  These features give a
measurement of the redshift of the reionization event.  The features are
reasonably sharp, with $\Delta\nu/\nu\approx 0.04$.  Haiman, Rees \& Loeb
(1997) estimate a similar effect whereby radiative excitations serve to
brighten Ly$\alpha$.  We find that recombinations give a significantly brighter
signal.  The recombination rate is proportional to the square of the gas
density, and thus is sensitive to the clumpiness of the gas, while the
excitation rate scales linearly with the gas density.  In a hierarchically
clustered universe we can expect that the recombination rate should be larger
than the excitation rate.

The positions of the broad maxima from the Lyman and Balmer series provide a
check on the redshift of the sharp absorption features.  These can also be used
to relate the redshift of the reionization event to the evolution of the
recombination rate.  These can provide valuable information about the evolution
of the ionizing background and the temperature evolution at the reionization
epoch.  These effects will be investigated in a future paper (Baltz, Gnedin, \&
Silk 1997).

\section{Discussion}

This signature is quite weak, though it may be feasible to see some of these
features in the near infrared, where dust emission is lowest.  Estimating the
broadband intensity $\nu J_\nu$ for spectra A-D at the position of the
Ly$\alpha$ feature, we get results of $1-5\times 10^{-3}$~ nW~ sr$^{-1}$~
m$^{-2}$.  The signal is currently below observational upper limits by a factor
of $\sim 10^4$ (Kashlinsky et al. 1996b), but it may not be completely obscured
according to theoretical models of the correlated component of the diffuse
light from young galaxies (Kashlinsky et al. 1996b, Veeraraghavan 1996).  All
of these predict that the broadband diffuse infrared light should be of the
order of $1-3$~ nW~ sr$^{-1}$~ m$^{-2}$.  The present model also predicts
correlations in the diffuse light.  In principle measuring these correlations
can distinguish between reionization by stars and reionization by quasars, as
the fluctuation length in the quasar scenario is much longer.

There are two avenues which seem to greatly increase our chances of detecting
these spectral features.  The first is a spaceborne mission such as the EGBIRT
or DESIRE proposals (Mather \& Beichmann 1996), which plan to put an infrared
telescope at 3 AU where the interplanetary dust scattering and emission is
reduced by a factor of at least 100.  Since such dust is the primary
foreground, such a mission would be quite fruitful.  The second involves
measuring the anisotropy of the CIB.  It has been shown that while foregrounds
are far brighter than any expected extragalactic infrared background, the
extragalactic sources contribute most of the fluctuations at the degree scale
down to the arcminute scale (Kashlinsky, Mather \& Odenwald 1996a, Kashlinsky
et al. 1996b).  If it were possible to measure fluctuations on scales smaller
than this, the recombinant emission should dominate due to its high redshift.
In fact, the angular size of the simulation boxes at the reionization events
are all on the order of a few seconds of arc.  For example, in runs B and C the
reionization event occurs at redshift $z=7.3$.  The comoving nonlinear scale at
this time is 100 $h^{-1}$ kpc.  This region subtends $4\farcs 5$ today.  Any
fluctuations at this scale should be produced at the reionization epoch, long
before the arcminute scale fluctuations characteristic of galaxy formation.

The spectra we have presented are only meant to be representative of what might
be observed.  There are large theoretical uncertainties concerning the
reionization epoch, and we hope a measurement like the one described in this
work can help to sort out the details of this important event in the history of
structure formation.

\acknowledgements

E. B. was supported by NASA grant 1-443839-23254-2.  N. G. was
supported by the UC Berkeley grant 1-443839-07427.  Simulations were performed
on the NCSA Power Challenge Array under the grant AST-960015N and on the NCSA
Origin2000 mini super-computer under the grant AST-970006N.

\newpage
\begin{figure}
\plotone{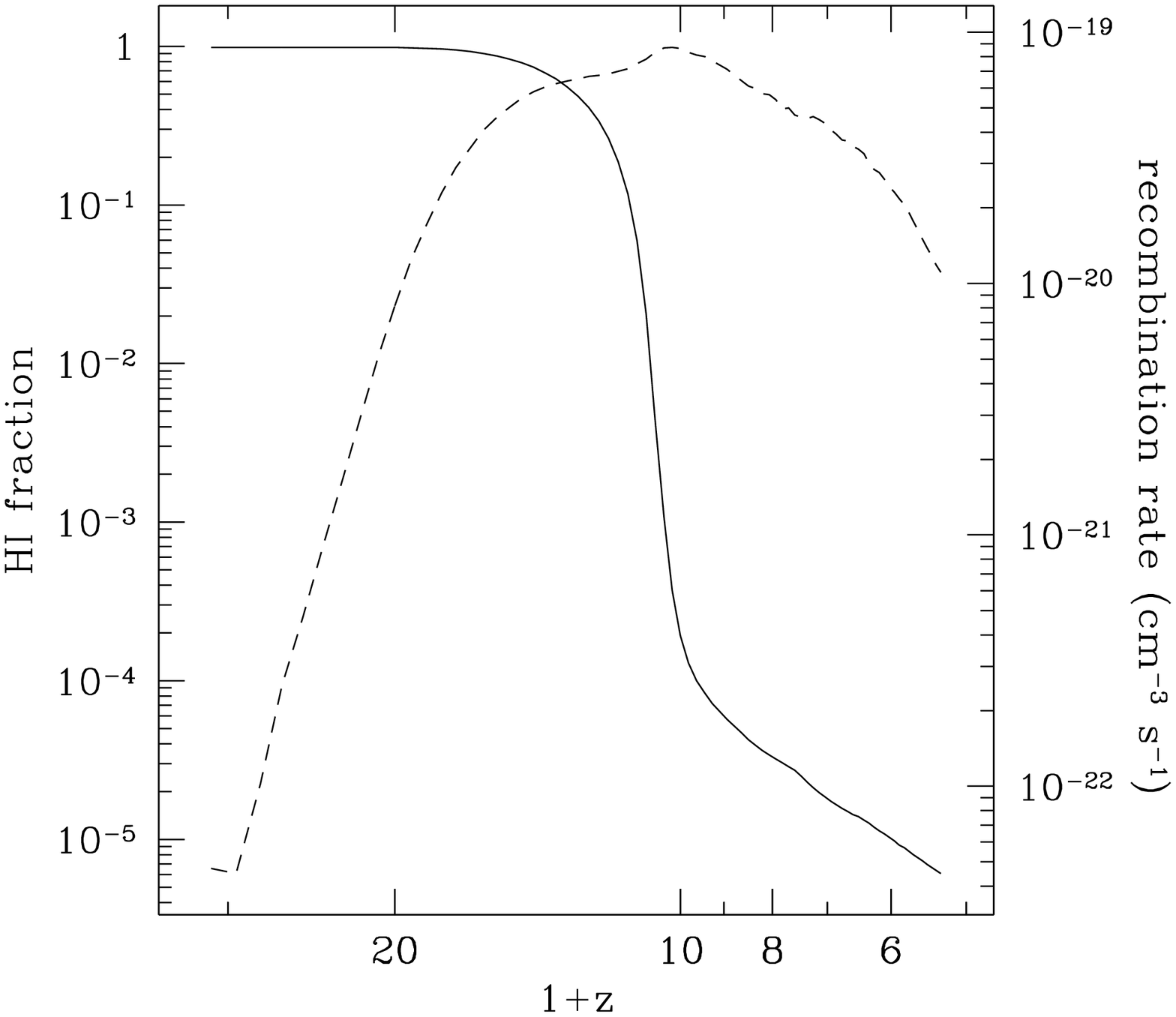}
\caption{Evolution of the recombination rate (dashed line) and
the neutral hydrogen fraction (solid line) in simulation A.  Clearly
reionization occurs rapidly, but the peak in recombinant emission rate is
quite broad.}
\end{figure}

\begin{figure}
\plotone{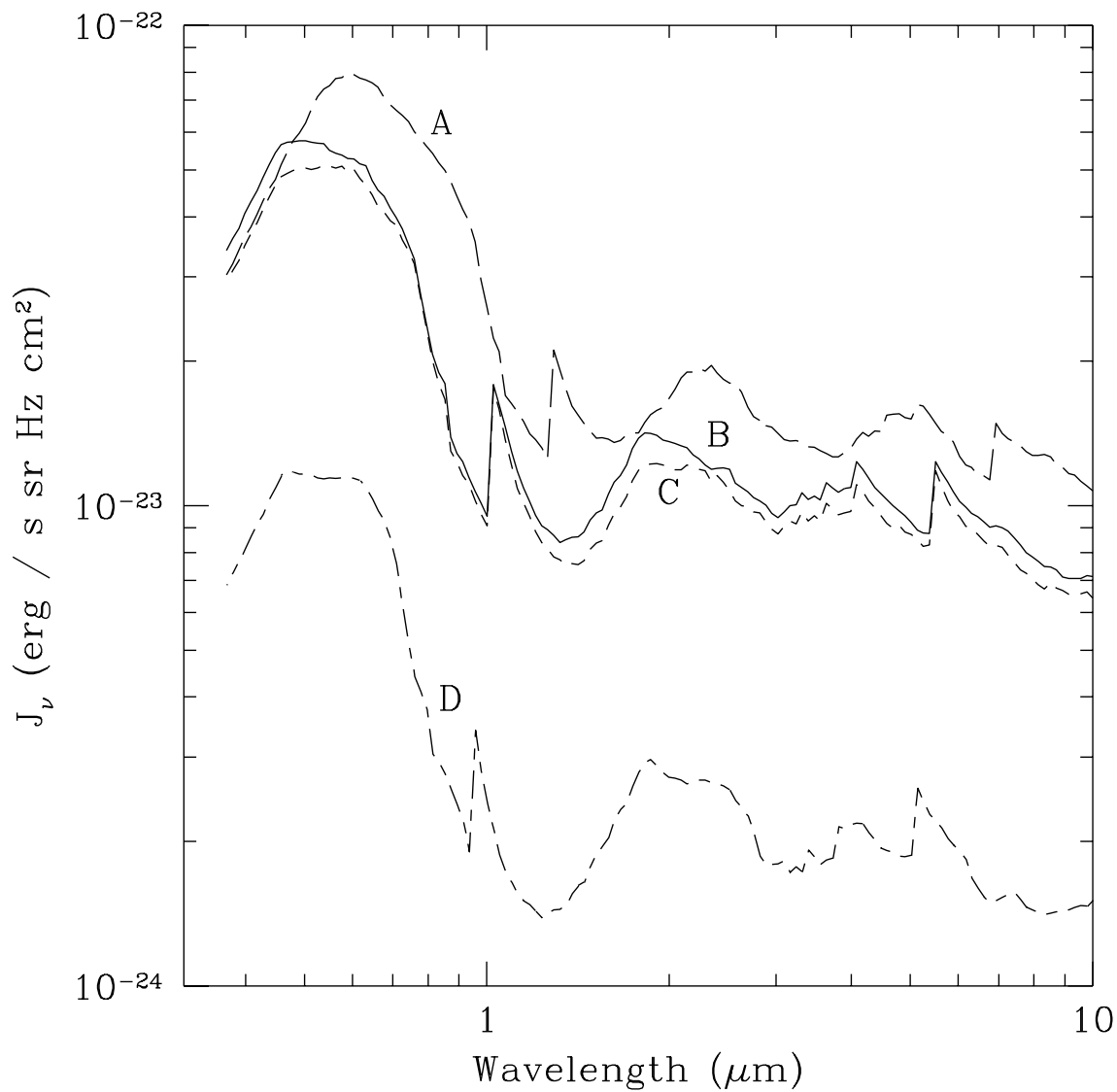}
\caption{Spectra for models A-D.  The recombination rate is
extrapolated to $z=0$ based on simulations E and F.  The brightening of both
Ly$\alpha$ at 1$\mu$m and H$\alpha$ at 6$\mu$m is clearly seen. Run A
undergoes reionization slightly earlier because its higher resolution allows
star formation to occur at a higher redshift.}
\end{figure}

\newpage
\tablenumpar

\end{document}